\documentclass[twocolumn,aps,prl]{revtex4-1}

\usepackage{slashed}
\usepackage{tikz-cd}
	\usepackage{amsmath}
	\usepackage{amsfonts}
	\usepackage{amssymb}
	\usepackage{graphicx}
	\usepackage{mathtools}
	\usepackage{stmaryrd}
	\usepackage[colorlinks=true,citecolor=blue,linkcolor=red]{hyperref}
	\usepackage{bbold}					
	\usepackage[makeroom]{cancel}		
	\usepackage{multirow}				
	\usepackage[normalem]{ulem}        
	\usepackage{array}

	\newcolumntype{x}[1]{>{\centering\let\newline\\\arraybackslash\hspace{0pt}}p{#1}}
	
	\makeatletter
\newcommand*\rel@kern[1]{\kern#1\dimexpr\macc@kerna}
\newcommand*\widebar[1]{%
  \begingroup
  \def\mathaccent##1##2{%
    \rel@kern{0.8}%
    \overline{\rel@kern{-0.8}\macc@nucleus\rel@kern{0.2}}%
    \rel@kern{-0.2}%
  }%
  \macc@depth\@ne
  \let\math@bgroup\@empty \let\math@egroup\macc@set@skewchar
  \mathsurround\z@ \frozen@everymath{\mathgroup\macc@group\relax}%
  \macc@set@skewchar\relax
  \let\mathaccentV\macc@nested@a
  \macc@nested@a\relax111{#1}%
  \endgroup
}
\makeatother

	\DeclareMathOperator{\tr}{tr}  		

	\DeclareMathAlphabet{\mathbbold}{U}{bbold}{m}{n}
	
	\def\bQ{{\bf{Q}}}
	
	\def\bx{{\bf{x}}}
	
	\def\bk{{\bf{k}}}
	
	\def\bb{{\bf{b}}}

	\def\PRLgreater{\,{>}\,}
	\def\PRLless{\,{<}\,}
	\def\PRLequal{\,{=}\,}
	\def\PRLminus{\,{-}\,}
	\def\PRLplus{\,{+}\,}

	\newcounter{subeqn} %
	\makeatletter
	\@addtoreset{subeqn}{equation}
	\makeatother

\definecolor{XQ}{rgb}{1,0,0}
\definecolor{XQ}{rgb}{0,0,0} 

\newcommand{\beginsupplement}{%
        \setcounter{table}{0}
        \renewcommand{\thetable}{S\arabic{table}}%
        \setcounter{figure}{0}
        \renewcommand{\thefigure}{S\arabic{figure}}
        \setcounter{equation}{0}
        \renewcommand{\theequation}{S\arabic{equation}}%
     }

\begin{document}
\title{Geometric Response and Disclination-Induced Skin Effects in Non-Hermitian Systems}
\author{Xiao-Qi Sun}\thanks{These two authors contributed equally.}
\author{Penghao Zhu}\thanks{These two authors contributed equally.}
\author{Taylor L. Hughes}
\affiliation{Department of Physics and Institute for Condensed Matter Theory,
University of Illinois at Urbana-Champaign, Urbana, Illinois 61801, USA}
\date{\today}

\begin{abstract}
We study the geometric response of three-dimensional non-Hermitian crystalline systems with nontrivial point-gap topology. For systems with fourfold rotation symmetry, we show that in the presence of disclination lines with a total Frank angle which is an integer multiple of $2\pi$, there can be nontrivial one-dimensional point-gap topology along the direction of the disclination lines. This results in disclination-induced non-Hermitian skin effects. By doubling a non-Hermitian Hamiltonian to a Hermitian three-dimensional chiral topological insulator, we show that the disclination-induced skin modes are zero modes of the effective surface Dirac fermion(s) in the presence of a pseudomagnetic flux induced by disclinations. Furthermore, we find that our results have a field theoretic description, and the corresponding geometric response actions (e.g., the Euclidean Wen-Zee action) enrich the topological field theory of non-Hermitian systems. 
\end{abstract}

\maketitle

\emph{Introduction.--}Non-Hermitian Hamiltonians provide a natural formalism to describe wave phenomena in the presence of loss and gain, which are ubiquitous in both classical~\cite{Schomerus:2013,Zhen:2015,Longhi:2015,Zeuner:2015,Lu:2015,Poli:2015,Weimann:2016,Takata:2018,Zhou:2018,Cerjan:2018a,Ghatak:2020,Weidemann:2020,Helbig:2020,Hofmann:2020,Wang:2020,Palacios:2020} and quantum~\cite{Choi:2010,Xiao:2017,Zhan:2017,Kozii:2017,Xu:2017,Shen:2018,Papaj:2018,Zyuzin:2018,Xiao:2020,Nagai:2020} systems. 
Recently, there has been a growing interest in the interplay between non-Hermiticity and topology. The synergy of these two concepts has produced fruitful results in non-Hermitian crytalline systems, such as new transport and dynamical features~\cite{Rudner:2009,McDonald:2018,Song:2019a,Lee:2019topological,Wanjura:2020,Xue:2020,Li:2020topology,Schomerus:2020,Yi:2020,Lee:2020,Lee:2020a,Para:2020}, new forms of bulk-boundary correspondence~\cite{Hatano:1996,Lee:2016,Leykam:2017,Yao:2018,Yao:2018a,Alvarez:2018,Kunst:2018,Lee:2018,Xiong:2018,Yokomizo:2019,Zirnstein:2019,Song:2019,Imura:2019,Jin:2019,Yang:2020}, and non-Hermitian analogy of topological insulators~\cite{Shen:2018a,Hirsbrunner:2019,Xi:2019,Lieu:2020,Tonielli:2020,Altland:2020} and semimetals~\cite{Cerjan:2018,Carlstrom:2018,Okugawa:2019,Carlstrom:2019,Moors:2019,Yoshida:2019,McClarty:2019,Kawabata:2019,Budich:2019,Yang:2019,Wang:2019,Yoshida:2020,Sun:2020,Yang:2020a,Yang:2020fermion,Hu:2021}.

One of the most remarkable consequences of non-Hermiticity is the new class of topological systems~\cite{Gong:2018,Zhou:2019,Bergholtz:2019b,Kawabata:2019symmetry,Liu:2019b,Liu:2019a,Wojcik:2020,Denner:2020} without Hermitian analogs. These intrinsically non-Hermitian topological systems have topological invariants associated with a point gap~\cite{Gong:2018} at a reference energy in the complex energy plane.
In one spatial dimension, the nontrivial point-gap topology 
produces the celebrated non-Hermitian skin effect (NHSE)~\cite{Hatano:1996,Yao:2018a}, which generates an extensive number of states localized at the boundaries of a system~\cite{Lee:2019,Lee:2019hybrid,Liu:2019second,Okuma:2020,Zhang:2020,Borgnia:2020,Kawabata:2020higher,Okugawa:2020,Ma:2020,Fu:2021}. Recently, the magnetic-field-induced NHSE in three-dimensional (3D) non-Hermitian Weyl semimetals has been studied, which originates from a 3D nontrivial point-gap topology~\cite{Bessho:2020}. In addition to electromagnetic response, there have been extensive studies of the \emph{geometric} response of Hermitian topological systems both in the continuum limit \cite{Avron:1995,Wen:1996,Read:2009,Hughes:2011t,Bradlyn:2012,Hughes:2013,Abanov:2014,Gromov:2014,Gromov:2015,Biswas:2016} and at lattice level~\cite{Teo:2013,Benalcazar:2014,Shapourian:2015,Li:2020,Rao:2020,Peterson:2021,Liu:2021,Liu:2019}. However, the understanding of the interplay between geometry and non-Hermitian point-gap topology is still preliminary.

In this Letter, we consider the geometric response of 3D non-Hermitian crystalline systems having nontrivial point-gap topology. We show that disclination lines in rotationally invariant systems can support one-dimensional (1D) point-gap topology along the direction of the disclination lines, hence leading to a corresponding NHSE. By mapping the non-Hermitian problem to a 3D chiral topological insulator, we show that the disclination skin modes are zero modes of the effective surface Dirac fermions subjected to a pseudomagnetic field induced by the curvature singularities at the disclinations. Furthermore, we show that we can describe this phenomenon with the inclusion of geometric terms in the non-Hermitian field theory approach \cite{Kawabata:2020}.

\emph{Point-gap topology and NHSE.--} We shall first review point-gap topology and its relation to the NHSE in 1D and the magnetic-field-induced NHSE in 3D. For a non-Hermitian Bloch Hamiltonian $H$, one can define a point gap if its spectrum does not cross a reference energy $E\in\mathbb{C}$, i.e., $\det(H\PRLminus E)\neq 0$. This means that $\det(H\PRLminus E)$ is a non-zero complex number and can have a winding number in the 1D Brillouin zone (BZ) protected by the point gap:
\begin{equation}
\label{eq: W1}
    W_{1}(E)=-\int_{0}^{2\pi }\frac{dk}{2\pi}\frac{\partial}{\partial k}\arg\left[ \det\left(H(k)-E\mathbb{1}\right)\right].
\end{equation}
$W_{1}(E)$ has been shown to count the number of eigenstates at energy $E$ that are localized at a semi-infinite boundary~\cite{Okuma:2020}. Interestingly, $W_1(E)$ can be nonzero for a continuous region of energy $E$ in the complex energy plane (bounded by gap-closing points), and therefore indicates an extensive number of eigenstates localized at the boundary.  This phenomenon is known as the NHSE. Furthermore, for a Hamiltonian with a nontrivial $W_1(E)$, the long-lived excitations are chiral, and feature anomalous dynamics absent in Hermitian systems~\cite{Lee:2019topological}. To illustrate this property, one simple example is a one-band model $H(k)\PRLequal e^{-ik}$ with nontrivial winding $+1$ for each reference energy inside the unit circle. The most amplified mode is at $k\PRLequal\PRLminus\pi/2$ with chirality $\chi\PRLequal\operatorname{sgn}\left[\text{Re}\left(\partial H/\partial k\right)\right]\PRLequal\PRLplus 1$. The NHSE and anomalous dynamics are crucial features of 1D point-gap topology.

Similar to $W_1(E)$, one can define a quantized winding number $W_3(E)$ for a Bloch Hamiltonian $H(\bk)$ in the 3D BZ at a reference energy $E$
\begin{equation}
\label{eq: W3}
\begin{aligned}
   &W_3(E)=-\int_{\text{BZ}} \frac{d^3k}{24\pi^3}\epsilon^{ijk}\tr\Big{[}(\widetilde{H}^{-1}\partial_{k_i}\widetilde{H})
    \\
    &\times(\widetilde{H}^{-1}\partial_{k_j}\widetilde{H})
    (\widetilde{H}^{-1}\partial_{k_k}\widetilde{H}) \Big{]}, 
\end{aligned}
\end{equation}
where $\widetilde{H}(\bk)\equiv H(\bk)-\mathbb{1} E$. The generalization of the anomalous dynamics is straightforward: the long-lived excitation is a 3D Weyl fermion. In a uniform magnetic field the Weyl node exhibits chiral 1D Landau levels dispersing along the field direction. These anomalous 1D chiral modes can be shown~\cite{Bessho:2020} to be precisely 
ascribed to the nontrivial 1D winding $W_{1}(E)=W_{3}(E)\Phi/2\pi$ with $\Phi$ being the total magnetic flux, and hence a magnetic-field-induced NHSE. 

\begin{figure}[h]
\centering
\includegraphics[width=1\columnwidth]{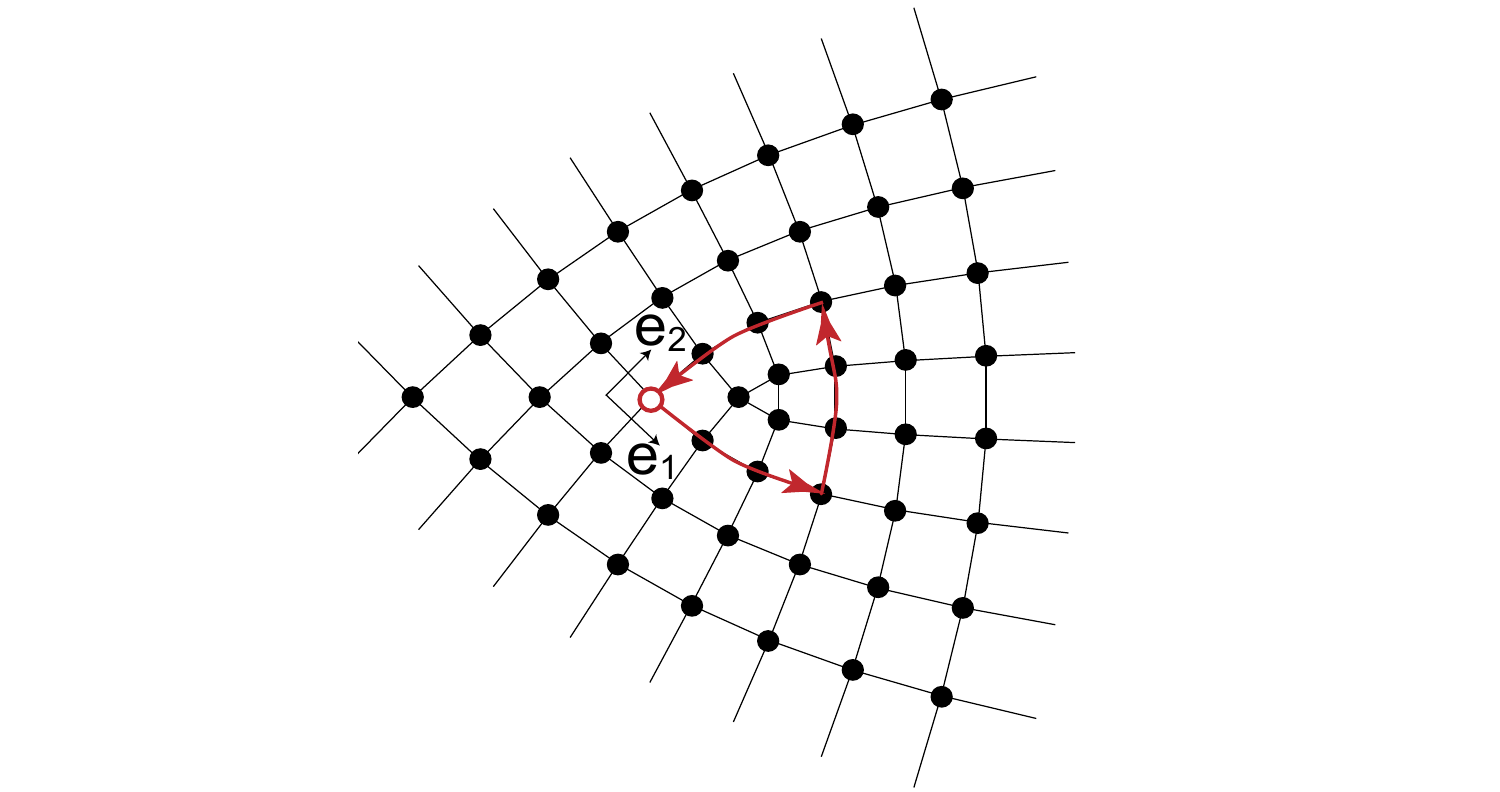}
\caption{$C_{4}$-symmetric lattice with a disclination. Starting from the red circle, the parallel transport around the loop provides net rotation of $-\pi/2$ and net transportation of $-3\mathbf{e}_{1}$. The Frank angle of this disclination is $\pi/2$. The equivalence class of Burgers vector is the parity of the sum of the Burgers vector components: $-3 \bmod 2=1$.
}
\label{fig:holonomy}
\end{figure}

\emph{Disclination-induced NHSE.--} Beyond the NHSE associated with electromagnetic fields, we can explore geometric probes of intrinsically non-Hermitian topology. We focus on lattice disclinations that heuristically serve as sources of geometric curvature.  A disclination is classified by a Frank angle and a Burgers vector equivalence class, which capture the amount of rotation and translation accumulated by a vector parallel transported along a loop encircling the disclination respectively. An example is given in Fig.~\ref{fig:holonomy}, and comprehensive discussions about the classification can be found in Refs.~\cite{Benalcazar:2014,Li:2020}. Due to the Frank angle and the Burgers vector (equivalence class), a wave packet will generically obtain a Berry phase when it adiabatically encircles the disclination, and thus the disclinations can effectively generate a pseudomagnetic field.  In the following, we will demonstrate that the pseudomagnetic flux induced by disclinations can lead to a NHSE in systems having nontrivial $W_3(E)$, and this effect is captured by a nontrivial effective 1D winding number $W_1(E)$ along the direction of the disclination line. 

\emph{Hermitian description.--} In order to see how the pseudomagnetic flux induced by disclinations leads to a NHSE in lattice models, it is convenient to introduce a doubled and Hermitianized Hamiltonian with chiral symmetry~\cite{Feinberg:1997, Gong:2018}:
\begin{equation}
\label{eq: hermitian}
\mathcal{H}(\bk)=
\left(
\begin{array}{cc}
   0  & H(\bk)^{\dagger}-E^{*} \\
 H(\bk)-E    & 0
\end{array}
\right),
\end{equation}
where $H(\bk)$ is the non-Hermitian Hamiltonian we want to study, and $E$ is the reference energy on which we focus.
There are two main features of this approach: (i) The topological winding number $W_3(E)$ of $H(\bk)$ equals the chiral winding number of $\mathcal{H}(\bk)$. Hence, the bulk-boundary correspondence of (Hermitian) chiral symmetric insulators indicates that a nonzero $W_{3}(E)$ implies the existence of $|W_3(E)|$ protected surface Dirac cones (SDCs). (ii) The existence of an \emph{exact} zero mode of $\mathcal{H}$ implies the existence of an eigenstate at energy $E$ ($E^{*}$) for $H$ ($H^{\dagger}$) depending on its chirality.
\begin{figure}[h]
\centering
\includegraphics[width=1\columnwidth]{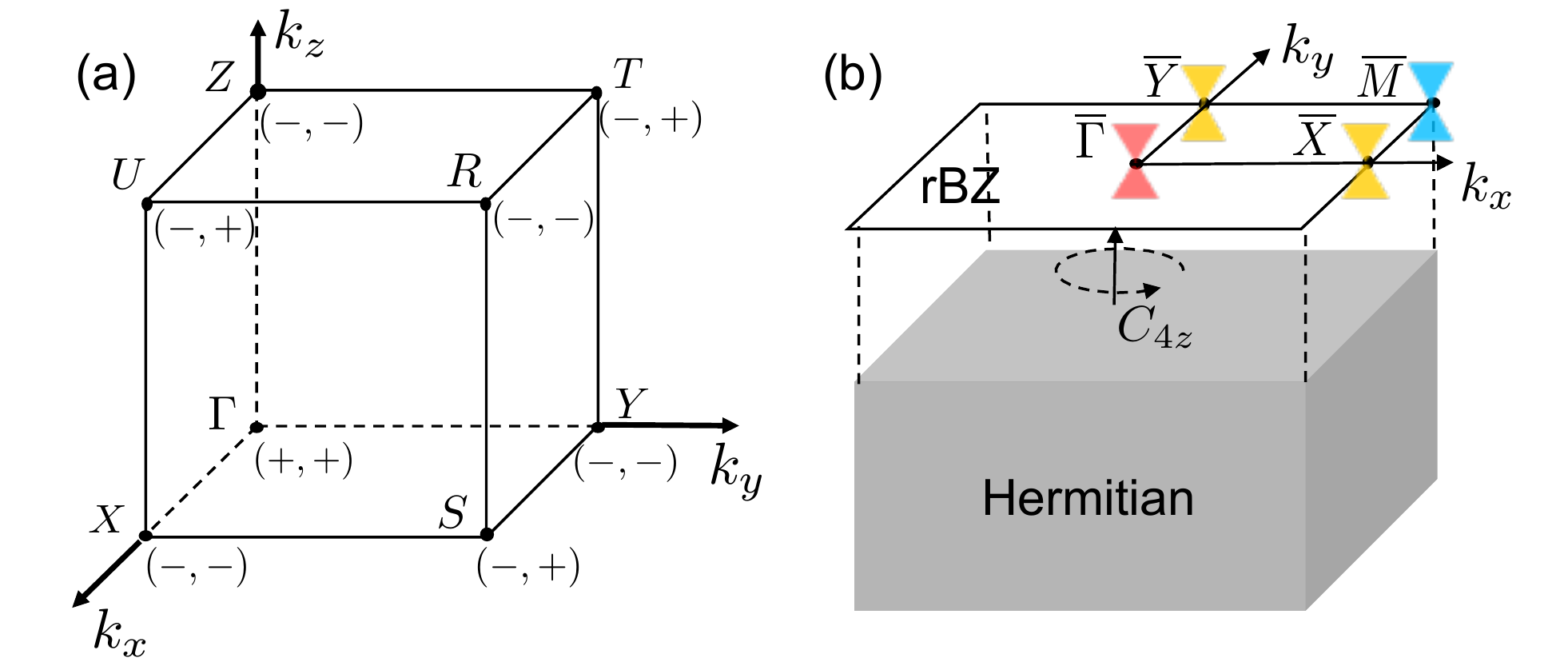}
\caption{(a) A tabulation of $(\operatorname{sgn}\left[\gamma-\text{Im}E\right],\chi_i)$ at the eight Weyl points of Eq.~\eqref{eq:nhw} when $1<\operatorname{Im}E<3$. (b) the locations of Dirac cone(s) on the surface of the Hermitian Hamiltonian in Eq.~\eqref{eq: hermitian} when $1<\operatorname{Im}E<3$ (red), $-1<\operatorname{Im}E<1$ (yellow), and $-3<\operatorname{Im}E<-1$ (blue). }
\label{fig:illustration}
\end{figure}

For an explicit illustration, let us focus on a concrete non-Hermitian Weyl semimetal model with a Bloch Hamiltonian: 
\begin{equation}
\label{eq:nhw}
    H(\bk)=t\sin k_x\sigma_x+t\sin k_y \sigma_y+t\sin k_z\sigma_z+i\gamma(\bk),
\end{equation}
where $\gamma(\bk)=\cos k_x+\cos k_y+\cos k_z$. This model has a $C_{4z}$ rotational symmetry represented by:
\begin{equation}
\label{eq:C4}
\begin{aligned}
    &C_{4z}=e^{-i\frac{\pi}{4}\sigma_z}e^{-i\pi \mathcal{L}_{z}/2},
    \\ 
    &C_{4z} H(k_x,k_y,k_z) C_{4z}^{\dagger}=H(k_y,-k_x,k_z),
\end{aligned}
\end{equation}
where the Pauli matrices represent the spin degree of freedom, and $\mathcal{L}_{z}$ is the angular momentum of the orbital that we put on each site, which takes values $-1,0,1,2$  $\operatorname{mod} 4,$ for $C_{4z}$-symmetric systems. We will see below that the orbital rotation phase in Eq.~\eqref{eq:C4} is important for the geometric response. For this model, the 3D winding number $W_3(E)$ has a simplified formula~\cite{Sun:2018b,Higashikawa:2019,Bessho:2020}:
\begin{equation}
\label{eq: W3_simple}
    W_3(E)=\sum_i \frac{1}{2}\text{sgn}\left[\gamma(\bQ_i)-\text{Im}E\right]\chi_i,
\end{equation}
where $\bQ_i$ are the momenta of the eight Weyl points in the Hermitian limit, and $\chi_i\PRLequal\pm 1$ are their corresponding chiralities. The positions of the Weyl points in momentum space are shown in Fig.~\ref{fig:illustration} (a), as well as their corresponding $(\operatorname{sgn}\left[\gamma-\text{Im}E\right],\chi_i)$. This model therefore has a nontrivial 3D point gap winding number of $W_3(E)\PRLequal1$ for $1\PRLless|\operatorname{Im} E|\PRLless3$, and $W_3(E)\PRLequal-2$ for $|\operatorname{Im} E|\PRLless1$.

Now we consider (the Hermitian) $\mathcal{H}$ in a semi-infinite bulk geometry terminated with a surface normal $\hat{z}$ and its two-dimensional (2D) reduced Brillouin zone (rBZ) [see Fig.~\ref{fig:illustration}(b)]. We start with the simple case where $1\PRLless\operatorname{Im} E\PRLless3$ and $W_3(E)\PRLequal1$. The low-energy effective Hamiltonian on the top $C_{4z}$-symmetric surface termination is a rotationally symmetric SDC at $\widebar{\Gamma}$ as shown in Fig.~\ref{fig:illustration}(b) :
\begin{equation}
\label{eq: conti}
    \mathcal{H}_{\text{eff}}=v (k_x\tau_x+ k_y\tau_y),
\end{equation}
where the Pauli matrices $\tau_i$ are the effective degrees of freedom, and the chiral symmetry is represented by $\tau_z$. 

Now we introduce disclination lines parallel to the $z$ direction, which correspond to curvature singularities in the continuous effective Dirac model [e.g. Eq.~\eqref{eq: conti}]. The curvature is represented by the coupling to a spin connection $\omega_i$ via the rotation generators. The Frank angle of disclination lines can be expressed in terms of $\omega_{i}$ as  $\Theta_z\PRLequal\int (\partial_x\omega_y\PRLminus\partial_y\omega_x)dx dy$.  In addition to the coupling to $\tau_z$, the spin connection also couples to $\mathcal{L}_z\mathbb{1}$, which adds an effective gauge flux $\mathcal{L}_{z}\Theta_{z}$ to the Dirac fermion~\cite{Supp}. According to the index theorem~\cite{nakahara:2003}, this flux $\mathcal{L}_{z}\Theta_{z}$, will lead to robust zero modes on the top surface with a total number of $\nu\PRLequal|\mathcal{L}_{z}\Theta_z/(2\pi)|$. Furthermore, the zero modes are eigenmodes of $\tau_z$ with eigenvalue (chirality) $\tau\PRLequal\text{sgn}~{\mathcal{L}_{z}\Theta_z}$~\cite{Supp}. For $\mathcal{L}_{z}\Theta_z\PRLgreater0$, the eigenmodes will have $\tau\PRLequal+1$ and correspond to skin modes of $H$ on the top surface and at energy $E$. By a similar argument for the bottom surface, one can derive that for $\mathcal{L}_{z}\Theta_z\PRLless0$, there are zero modes of $\mathcal{H}$ corresponding to skin modes of $H$ on the bottom surface and at energy $E$. It is known~\cite{Okuma:2020} that the number of zero modes of $\mathcal{H}$ with $\tau\PRLequal+1$  equals $|W_{1}(E)|$, and that $\text{sgn}[W_{1}(E)]\PRLequal+/-$ indicates that the zero modes with $\tau\PRLequal+1$ are on the top/bottom surface. Hence, we can conclude that when $1\PRLless\operatorname{Im} E<3$ and $W_3(E)\PRLequal1$, the disclination induces a winding number of $W_{1}(E)\PRLequal\mathcal{L}_{z}\Theta_z/(2\pi)$. 
We note that the effective flux is contributed \emph{only} by the orbital rotation generator, which commutes with $\tau_i$. From here on, we focus only on this part to compute the effective flux for a Dirac fermion in order to determine $W_1(E)$.

Let us proceed to discuss cases where the SDCs are at momenta away from $\widebar{\Gamma}$. Since disclinations are classified by a Frank angle $\Theta_{z}$ and a Burgers vector (equivalence class) of $\bb\PRLequal(b_{x},b_{y})$, for a SDC with nonzero momentum $\bQ_{\perp}$, the Burgers vector will also contribute an effective flux of $-\bQ_{\perp}\cdot \bb$. With this consideration in mind, we analyze other topologically nontrivial regimes of our model. First, when $-3\PRLless\operatorname{Im}E\,{<}-1$, there is a single SDC at $\widebar{M}$ [see Fig.~\ref{fig:illustration} (b)], and the effective flux it feels is~\cite{Supp} 
\begin{equation}
\label{eq: efluxM}
\Phi^{\text{eff}}(\widebar{M})=\mathcal{L}_{z}\Theta_{z}-(\pi,\pi)\cdot \bb.
\end{equation}
From the index theorem for 2D Dirac fermions, as well as the relation between $W_{1}$ and the number of surface zero modes, it is straightforward to see that the effective flux in Eq.~\eqref{eq: efluxM} leads to a winding number of $W_{1}(E)\PRLequal\Phi^{\text{eff}}(\widebar{M})/(2\pi)$. Second, when $-1\PRLless\operatorname{Im}E\PRLless1$, there is a pair of SDCs at $\widebar{X}$ and $\widebar{Y}$ [see Fig.\ref{fig:illustration} (b)]. Since a $C_{4z}$ rotation takes one SDC to another, and the translation phase obtained by each SDC is different, these two SDCs together form an irreducible representation of the space group. A subtlety arises that we cannot \emph{individually} define the effective flux for each SDC, and we need to consider the (possibly \textit{non-Abelian}) effective flux of both. In doing so, we write the $C_{4z}$ rotation operator for these two SDCs acting on the \emph{orbital} degrees of freedom: \begin{equation}
\label{eq:c4generator}
    C_{4z}=\sigma_x^{v}e^{-i\frac{\pi}{2}\mathcal{L}_z}=\exp\left(-i\frac{\pi}{2}\mathcal{L}_z\sigma_x^{v}\right),
\end{equation}
where the superscript $v$ indicates the operator is in the valley space spanned by $\widebar{X}$ and $\widebar{Y}$, and $\sigma_{x}^{v}$ exchanges the two valleys. From Eq.~\eqref{eq:c4generator}, we can see $\mathcal{L}_{z}\sigma_{x}^{v}$ is the orbital rotation generator, which leads to a non-Abelian flux for SDCs at the two valleys. The effective flux can be written as $\Phi_{r}\PRLequal\mathcal{L}_{z}\Theta_{z}\sigma_{x}^{v}$, where $\Theta_{z}$ is a multiple of $\pi/2$ in the $C_{4z}$-symmetric case. In addition, the translation phase of SDCs at the two valleys contributes another matrix flux:
\begin{equation}
\label{eq: efluxXY}
\Phi_{t}=-\pi\begin{pmatrix} b_x &0 \\ 0 & b_{y} \end{pmatrix}.
\end{equation}
Combining these two contributions, we define the total non-Abelian effective flux felt by SDCs at two valleys to be 
\begin{equation}
e^{i\Phi^{\text{eff}}(\widebar{X}\widebar{Y})}=e^{i \Phi_{t}}e^{i\Phi_{r}}.
\end{equation}
Now applying the index theorem again~\cite{Supp}, we find a winding number of $W_1(E)\PRLequal-\text{tr}[\Phi^{\text{eff}}(\widebar{X}\widebar{Y})]/(2\pi)$, where the minus sign comes from the chiral winding number of $-1$ of the SDCs. Notice that the analysis from the index theorem is for the effective Dirac Hamiltonian in the continuum limit, however, we will show that this indeed captures the disclination-induced skin effect in our lattice model [cf. Eq.~\eqref{eq:nhw}].
\begin{figure}[h]
\centering
\includegraphics[width=1\columnwidth]{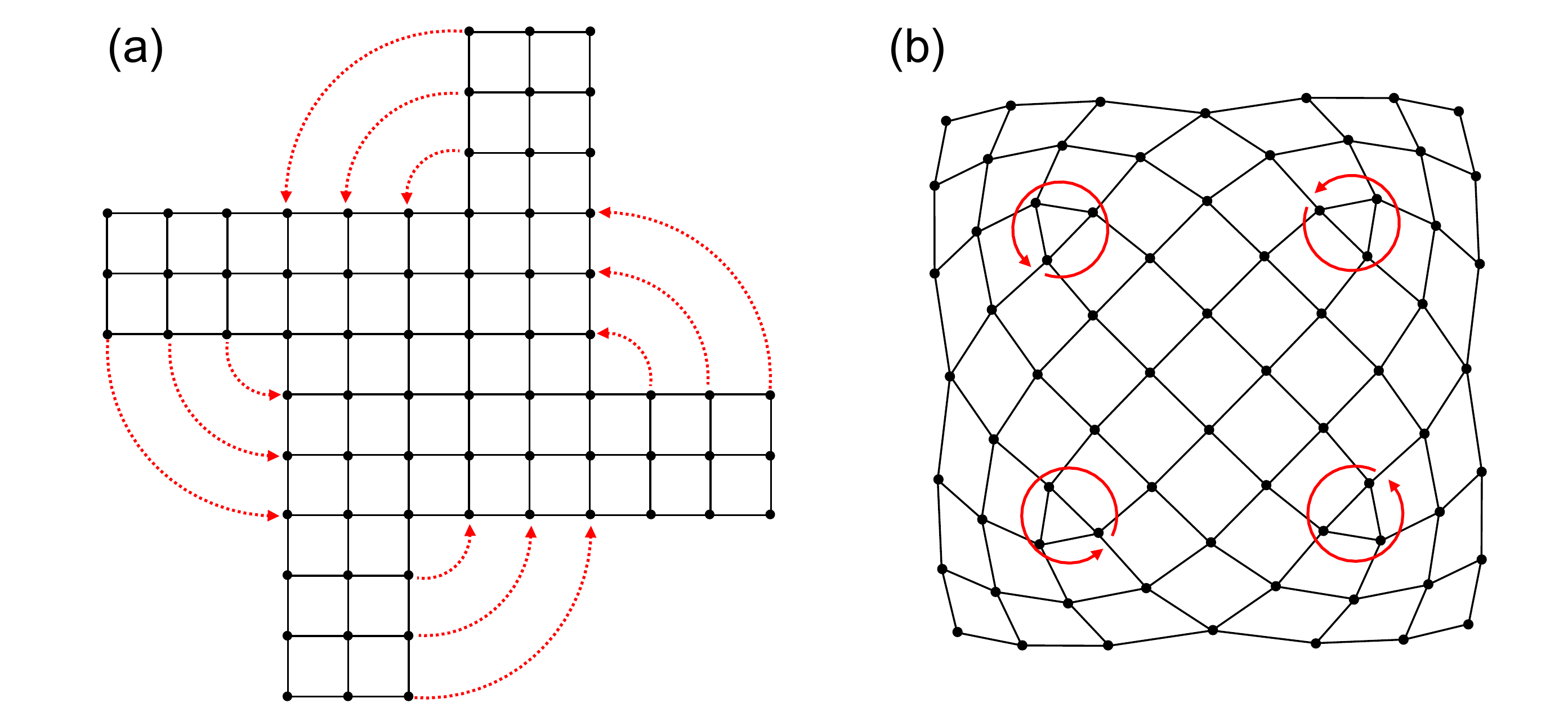}
\caption{Schematic illustration of a lattice construction with four disclinations, each of which has a Frank angle of $\pi/2$. (a) is the gluing procedure \cite{Supp}, and (b) is a completed lattice with disclinations. }
\label{fig:disclination}
\end{figure}

\emph{Numerical results.--}
To find a quantized $W_1(E)$ and the corresponding NHSE, we need a total pseudomagnetic flux that is time-reversal odd and an integer multiple of $2\pi$. However, there is a subtlety on the lattice that a flux of $n\pi$ passing through a single lattice plaquette is time-reversal even and incompatible with the chiral anomalous dynamics and the corresponding NHSE. Thus, for our lattice calculations this motivates us to consider orbital angular momentum $\mathcal{L}_z\PRLequal\pm 1$ and introduce four disclinations~\footnote{Here we use open boundary conditions in $x$ and $y$ directions. We also discuss the case of eight disclinations with total Frank angle $4\pi$ in the Supplemental Material~\cite{Supp}, which enables us to compactify $x$ and $y$ directions to a 2-sphere. The numerical results agree with our Hermitian description and suggest that the appearance of disclination-induced NHSE along $z$ direction does not depend on boundary physics in $x$ and $y$ directions in our model.}, each with a Frank angle of $\Theta^{s}_{z}\PRLequal\pm\pi/2$, where the superscript $s$ is used to distinguish the Frank angle of a \textit{single} disclination and the total Frank angle $\Theta_{z}$ in the system. For illustration purposes, let us fix $\mathcal{L}_z\PRLequal1,$ and focus on the case where each of the four disclinations has a Frank angle of $\Theta_{z}^{s}\PRLequal\pi/2$ in the following. We discuss other cases, e.g.,  $(\Theta_{z}^{s}\PRLequal\pi/2,\,\mathcal{L}_{z}\,{=}-1)$,  and $(\Theta_{z}^{s}\,{=}-\pi/2,\,\mathcal{L}_{z}\,{=}\pm 1)$, etc., in the Supplemental Material~\cite{Supp}. 

\begin{figure}[h]
\centering
\includegraphics[width=1\columnwidth]{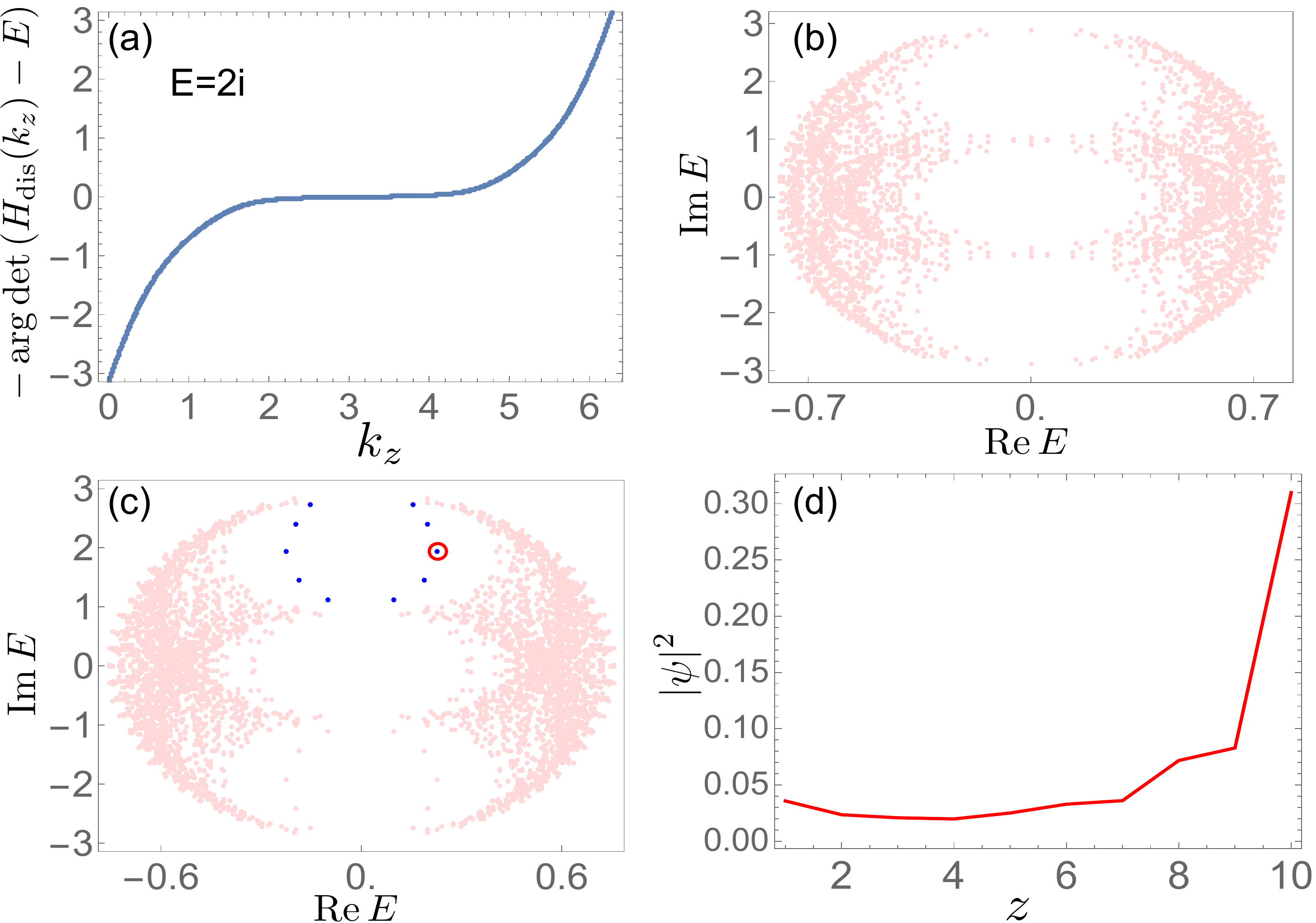}
\caption{Numerical calculations for the non-Hermitian Weyl semimetal described by Eq.~\eqref{eq:nhw} on a lattice with four disclinations as shown in Fig.~\ref{fig:disclination}, and $t=1/2$ is used in all calculations. The lattice has 10 unit cells along $z$ direction, and 200 unit cells in the $x$-$y$ plane at each $z$. (a) shows the nontrivial one-dimensional point gap winding number $W_1(E)$ at $E=2 i$. (b) and (c) show the energy spectra under periodic and open boundary condition along $z$ direction, where the blue dots in (c) are skin modes on the 
top surface. (d) shows the wave function along $z$ direction for the state indicated by a red circle in (c).}
\label{fig:windingskin}
\end{figure}

For our numerical lattice calculations we construct four disclinations with a Frank angle of $\Theta_{z}^{s}\PRLequal\pi/2$ by a process shown in Fig.~\ref{fig:disclination}.
Then, we add the hopping terms determined by the Hamiltonian in Eq.~\eqref{eq:nhw} on the disclinated lattice to derive a Hamiltonian $H_{\text{dis}}(k_{z})$. We note that we are using disclinations with ``plaquette-type" cores \cite{Teo:2013}, which implies a Burgers vector \emph{class} ${\bf{b}}^{s}\PRLequal a\hat{x},$ where $a$ is the lattice constant and the superscript $s$ indicates the Burgers vector is for a \textit{single} disclination. Note that because of the $C_{4z}$ symmetry we could equivalently say that ${\bf{b}}^{s}\PRLequal a\hat{y},$ hence a Burgers vector class. Let us first focus on the simplest regime when $1\PRLless\operatorname{Im}E\PRLless3$ and $W_{3}(E)=1$. To show the nontrivial 1D point gap winding number $W_{1}(E)=1$ [cf. Eq.~\eqref{eq: W1}] of $H_{\text{dis}}(k_{z})$,
we plot $\PRLminus\operatorname{arg}\det\left(H_{\text{dis}}(k_z)-E\mathbb{1}\right)$ at $E=2i$ in Fig.~\ref{fig:windingskin}(a). Figure~\ref{fig:windingskin}(b) shows the spectrum of $H_{\text{dis}}$ under periodic boundary conditions along the $z$ direction, where a loop circling the  $1\PRLless\operatorname{Im} E\PRLless3$ region can be seen. The skin modes on the \textit{top} surface, which are qualitatively captured by $W_{1}(E)\PRLequal1,$ are indicated by blue dots in the spectrum of open boundary conditions in the $z$ direction, as shown in Fig.~\ref{fig:windingskin}(c). We find ten skin modes in the region of $1\PRLless\operatorname{Im}E\PRLless3,$ which is consistent with the extensive NHSE for a 1D line in a 3D system where $N_z\PRLequal10$. In Fig.~\ref{fig:windingskin}(d), we show an exponentially decaying wave function of a representative skin mode [circled in Fig.~\ref{fig:windingskin}(c)]. Hence, when $1\PRLless\operatorname{Im}E\PRLless3$, $\mathcal{L}_{z}\PRLequal\PRLplus1$, and there is pseudomagnetic flux $2\pi$, we have shown numerically that $W_{1}(E)\PRLequal1$ and there are corresponding skin modes, which is consistent with our Hermitian description above \footnote{Numerical results for all other cases with $1\PRLless\operatorname{Im}E\PRLless3$ (\textit{i.e.},  $\mathcal{L}_{z}\PRLequal\PRLminus 1$ and pseudomagnetic flux $2\pi$, and $\mathcal{L}_{z}=\pm 1$ and pseudomagnetic flux $\PRLminus 2\pi$) are shown in the Supplemental Material, and are also consistent with our Hermitian description. 
}.  

\begin{figure}[h]
\centering
\includegraphics[width=1\columnwidth]{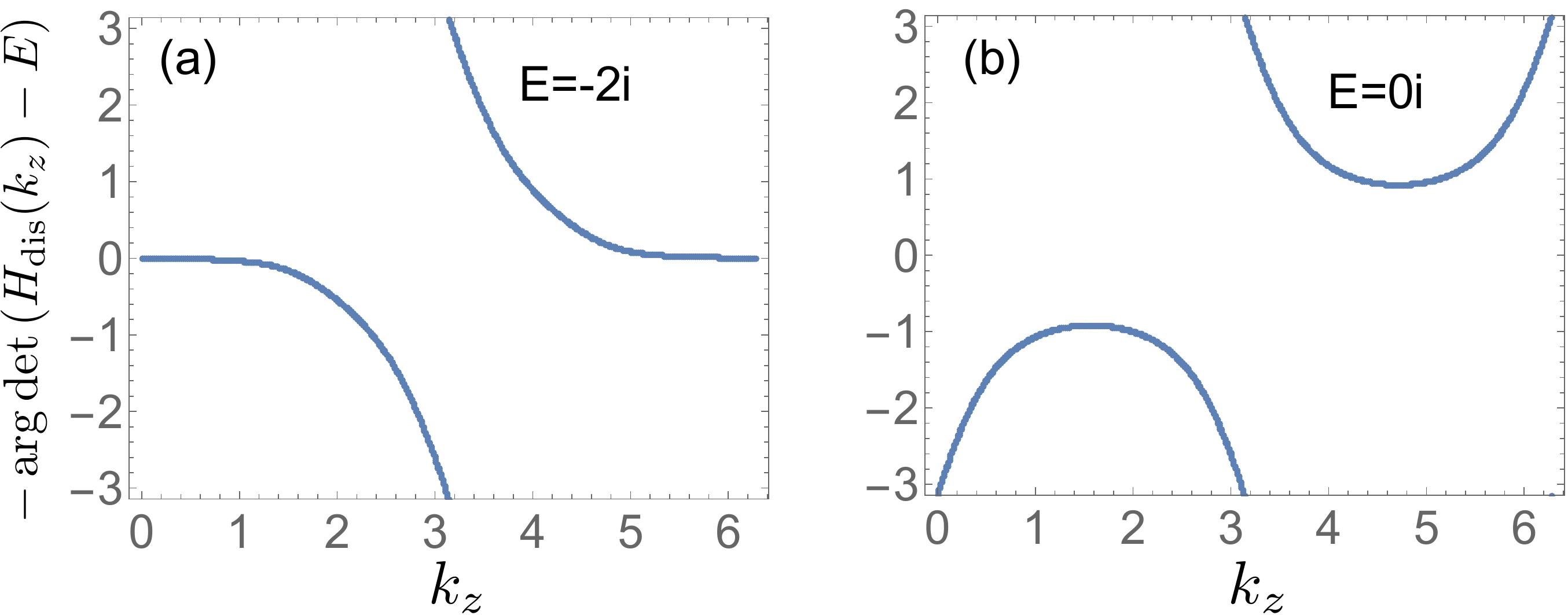}
\caption{The 1D point-gap winding number $W_1(E)$ on a lattice with four disclinations as shown in Fig.~\ref{fig:disclination} for (a) $E=-2i$ and (b) $E=0i$.} 
\label{fig:caveat}
\end{figure}

Next, consider the regime when $-3\PRLless\operatorname{Im}E\,{<}-1$. 
Since each of the four disclinations in Fig.~\ref{fig:disclination} has Burgers vector class $\bb^{s}\PRLequal a\hat{x}$, the effective flux [cf. Eq.~\eqref{eq: efluxM}] contributed from each disclination is $-\pi/2$. In total, there is a $\Phi^{\text{eff}}(\widebar{M})\PRLequal-2\pi$ pseudomagnetic flux, and we expect a nontrivial 1D point-gap winding of $W_{1}(E)\PRLequal-1$, which is confirmed by our numerical calculation shown in Fig.~\ref{fig:caveat} (a). Finally, consider the regime when $-1\PRLless\operatorname{Im}E\PRLless 1$. By substituting $\bb^{s}\PRLequal a\hat{x}$ and $\Theta^{s}_{z}\PRLequal\pi/2$ into $\Phi_{t}$ and $\Phi_{r}$ defined above, one can calculate $e^{i\Phi^{\text{eff}}(\widebar{X}\widebar{Y})}\PRLequal\sigma_{y}^{v}$, of which the two eigenmodes feel time-reversal 
even flux ($\Phi_1\PRLequal 0$ and $\Phi_2\PRLequal\pi$) on each disclination, and thus we expect a trivial 1D point-gap winding of $W_{1}(E)\PRLequal0$. This is confirmed by our numerical calculation shown in Fig.~\ref{fig:caveat}(b).

\emph{Conclusions and discussions.--}We studied the geometric response of a $C_{4z}$-symmetric 3D non-Hermitian lattice model with nontrivial point-gap topology characterized by $W_3$. We found that disclinations can induce NHSEs along the disclination lines. Interestingly, we note that our results can be described by the framework of the Euclidean field theory for non-Hermitian systems recently proposed in Ref.~\cite{Kawabata:2020}. The disclination-induced NHSE can be captured by a Euclidean Wen-Zee term encoding the Frank angle contribution\cite{Wen:1996}:
\begin{equation}
  S_{\text{WZ}}=\frac{\mathcal{L}_{z} W_3(E)}{2\pi}\int d^3\bx \epsilon^{ijk}A_i\partial_j\omega_k,
   \label{WZ}
\end{equation}
and a term containing the mixing of the $U(1)$ gauge field and the frame/translation gauge field for the Burgers vector equivalence class. These geometric response actions enrich the types of topological field theory responses for non-Hermitian systems. Interested readers are referred to the Supplementary Material for more details \cite{Supp}.   

It is also important to note that our model is expected to be realizable in a variety of platforms~\cite{Schomerus:2013,Zhen:2015,Longhi:2015,Zeuner:2015,Lu:2015,Poli:2015,Weimann:2016,Takata:2018,Zhou:2018,Cerjan:2018a,Ghatak:2020,Weidemann:2020,Helbig:2020,Hofmann:2020,Wang:2020,Palacios:2020} previously used to study non-Hermitian physics. Besides directly observing the boundary localized skin modes, one can also probe the disclination-induced skin effect using the chiral propagation of wave packets created near disclinations in the bulk. Furthermore, our model study serves as a first concrete example to theoretically understand geometric response in non-Hermitian topological systems, which will motivate the pursuit of the non-Hermitian analogy of more geometric responses such as Hall viscosity in future research.

\begin{acknowledgments}
\emph{Acknowledgments.--} X.-Q.~S. acknowledges support from the Gordon and Betty Moore Foundations EPiQS Initiative through Grant GBMF8691. P.~Z. and T.~L.~H. thank the US Office of Naval Research (ONR) Multidisciplinary University Research Initiative (MURI) grant N00014-20- 1-2325 on Robust Photonic Materials with High-Order Topological Protection
for support.
\end{acknowledgments}

\bibliography{bib}{}
\bibliographystyle{apsrev4-1}

\onecolumngrid
\begin{widetext}
\section{Supplemental Material for: Geometric response and disclination-induced skin effects in non-Hermitian systems}

\beginsupplement

\section{Effective flux from index theorem}
In this section, we justify the index theorem for SDCs used in the main text. In our case, we have rotation (around $z$ direction) generator  $\mathcal{L}_{z}\mathbb{1}+\tau_{z}/2$ for a SDC (at $\widebar{\Gamma}$). When we introduce disclinations, the spin connection $\omega_{\mu}$ with $\mu=1,2$ will enter the Dirac operator, which can be written as
\begin{equation}
\label{eq: SDCD}
\slashed{D}_{SDC}=e^{\mu}_{a}\gamma^{a}\left[\mathbb{1}\partial_{\mu}-i \omega_{\mu}(\mathcal{L}_{z}\mathbb{1}+\tau_{z}/2)\right]=e^{\mu}_{a}\gamma^{a}(\mathbb{1}\partial_{\mu}-i \mathcal{L}_{z}\omega_{\mu}\mathbb{1}-i \omega_{\mu}\tau_{z}/2),
\end{equation}
where $e^{\mu}_{a}$ is the frame field and $\gamma^{1,2}=\tau_{x,y}$.
If we compare Eq.~\eqref{eq: SDCD} with the most general Dirac operator coupled to a U(1) gauge field and to the background geometry:
\begin{equation}
\label{eq:D}
\slashed{D}=e^{\mu}_{a}\gamma^{a}(\mathbb{1}\partial_{\mu}-i A_{\mu}\mathbb{1}-\frac{i}{4}\omega^{ab}_{\mu}\gamma_{ab}),
\end{equation}
where $\omega_{\mu}^{ab}$ is the spin connection. We can see $\omega_{\mu}$ in Eq.~\eqref{eq: SDCD} represents the spin connection component $\omega_{\mu}^{12}=-\omega^{21}_{\mu}$, and correspondingly $\gamma^{12}=-\gamma^{21}=\tau_{z}$. If we further consider $\mathcal{L}_{z}\omega_{\mu}$ as an effective Abelian gauge field $A^{\text{eff}}_{\mu}$, it is straightforward to see that Eq.~\eqref{eq: SDCD} is a special case of Eq.~\eqref{eq:D} on a manifold $M$ described by the fixed $\omega_{\mu}^{ab}$. Then we can apply the index theorem for the general Dirac operator (see Chapter 12 of Ref.~\cite{nakahara:2003}) to Eq.~\eqref{eq: SDCD}:
\begin{equation}
\label{eq:index}
\nu_{+}-\nu_{-}=\frac{1}{2\pi}\int_{M}\partial_{x}A^{\text{eff}}_{y}-\partial_{y}A^{\text{eff}}_{x}=\frac{\mathcal{L}_{z}}{2\pi}\int_{M}\partial_{x}\omega_{y}-\partial_{y}\omega_{x}=\frac{\mathcal{L}_{z}\Theta_{z}}{2\pi},
\end{equation}
where $\nu_{+}$ ($\nu_{-}$) is the number of zero modes with chirality $\tau=+1$ ($-1$). There are a total of $\nu\equiv|\nu_{+}-\nu_{-}|$ number of robust zero modes. In the case of  $\nu_{+}-\nu_{-}>0$, the $\nu$ robust zero modes have chirality $\tau=+1$, which corresponds to skin modes. We note that the RHS of Eq.~\eqref{eq:index} only has contribution from the flux of the (effective) abelian gauge field, and has no contribution from the coupling of spin connection to $\gamma_{ab}$, which is true for Dirac operators in 2D \cite{nakahara:2003}. We also note that the above discussion is for a SDC with chiral winding number $+1$, for SDCs with chiral winding number $-1$ there will be an extra minus sign on the RHS of Eq.~\eqref{eq:index}. To apply the index theorem in the main text, we should consider that the top SDC has chiral winding number $\text{sgn}[W_3(E)]$ and the bottom SDC have the opposite chiral winding number in our model. With similar analysis, we can also find the effective Abelian/non-Abelian gauge flux and use the correct index theorem in other cases. 

Before proceeding to other cases, we detour to show that the NHSE derived from the index theorem in the Hermitianized Hamiltonian discussed above also has a field theoretic description, which is of help to gain better understanding of the disclination induced NHSE. In Ref.~\cite{Kawabata:2020}, Kawabata \emph{et al.} proposes a Euclidean topological field theory for non-Hermitian systems with nontrivial point-gap topology. They formally introduce a Euclidean action coupling to a spatial gauge field for each reference energy $E$
\begin{equation}
    S_{E}[\boldsymbol{A},\psi_E]=\int  \psi_{E}^{\dagger}[H(-i\partial-\boldsymbol{A})-E]\psi_{E}d^{2n+1} x.
\end{equation}
Then, the effective action upon integrating out fermions is the $(2n+1)$-D Chern-Simons action with coefficient $W_{2n+1}(E)$.  For example, the Chern-Simons action in 1D:
\begin{equation}
S_{E}[A_{x}]=W_{1}(E)\int dx A_{x},
\end{equation}
leads to a chiral current
\begin{equation}
j_{x}=\frac{\delta S_{E}[A_{x}]}{\delta A_{x}}=W_{1}(E),
\end{equation}
which can capture the chiral anomalous dynamics and the NHSE in 1D. Similarly, the Chern-Simons action in 3D, $i.e.$,
\begin{equation}
\label{eq:CS3D}
S_{E}[\boldsymbol{A}]=\frac{W_{3}(E)}{4\pi}\int d^{3}\epsilon_{ijk}A_{i}\partial_{j}A_{k}
\end{equation}
can capture the magnetic field induced NHSE  by a dimension reduction:
\begin{equation}
\label{eq: dimred}
\frac{W_{3}(E)}{4\pi}\int d^{3}\epsilon_{ijk}A_{i}\partial_{j}A_{k}
=\frac{W_{3}(E)}{2\pi}\int_{S} B_{z} dxdy\int dz A_{z}\equiv W_{1,z}(E)\int dz A_{z} \ \Rightarrow \ W_{1,z}(E)=\frac{W_{3}(E)}{2\pi}\int_{S} B_{z} dxdy.
\end{equation}
Motivated by Eqs.~\eqref{eq:CS3D} and \eqref{eq: dimred}, it is straightforward to see the NHSE due to the winding number $\mathcal{L}_{z}\Theta_{z}/(2\pi)$ can be captured by a Euclidean Wen-Zee term:
\begin{equation}
  S_{\text{WZ}}=\frac{\mathcal{L}_{z} W_3(E)}{2\pi}\int d^3\bx \epsilon^{ijk}A_i\partial_j\omega_k.
   \label{WZ}
\end{equation}
When have a disclination line along $z$ direction, Eq.~\eqref{WZ} leads to
\begin{equation}
\label{eq: disW1}
     W_{1}(E)=\frac{\mathcal{L}_{z} W_{3}(E)}{2\pi}\int_{S} \Omega_z dxdy=\frac{\mathcal{L}_{z} W_{3}(E)\Theta_{z}}{2\pi},
\end{equation}
where $\Omega_z\equiv \partial_x\omega_y-\partial_y\omega_x$.

 With the field theoretic description in mind, let us come back to the discussion about index theorem in other cases. When the SDC at $\widebar{M}$, we add an extra flux related to translation, $-(\pi,\pi)\cdot\bb$, to the effective flux. This is because for SDCs with non-zero momenta $\bQ_{\perp}=(Q_{x},Q_{y})$, there will be another effective gauge field $-Q_{\perp,a}e_{\mu}^{a}$ besides $\omega_{\mu}\mathcal{L}_{z}$ and the extra effective gauge flux can be computed as
\begin{equation}
\label{eq: translationflux}
-Q_{\perp,a}\oint dx^{\mu} e^{a}_{\mu}=-Q_{\perp,a}b^{a}=-\bQ_{\perp}\cdot \bb.
\end{equation}
Like the effective flux from $\omega_{\mu}\mathcal{L}_{z}$ can be captured by a bulk Wen-Zee term in Eq.~\eqref{WZ}, we tentatively propose a term at low energy to capture the translation phase:
\begin{equation}
S_{t}=\frac{1}{2\pi}\int d^{3}\bx \epsilon^{ijk}A_{i}Q_{\perp,a}\partial_{j}e^{a}_{k},
\end{equation}
the details of which will be left to future research. 

When there are a pair of SDCs at $X$ and $Y$ of the same winding number, we have a 2D Dirac operator under a non-Abelian effective gauge field related to orbital angular momentum, $\mathcal{L}_{z}\sigma_{x}^{v}\otimes\mathbb{1}$, and a non-Abelian effective gauge field related to translation, $\begin{pmatrix}(\pi,0)\cdot \mathbf{e}_{\mu}& 0
\\
0 &(0,\pi)\cdot \mathbf{e}_{\mu}&\end{pmatrix}\otimes\mathbb{1}$. Then, we need to consider the non-Abelian effective flux $\Phi^{\text{eff}}(\widebar{X}\widebar{Y})$ contributed by both and we have 
\begin{equation}
    \nu_{+}-\nu_{-}=\pm\text{sgn}[{W_3(E)}]\frac{\text{tr}\left[\Phi^{\text{eff}}(\widebar{X}\widebar{Y})\right]}{2\pi},
\end{equation}
for the top/bottom surface as used in the main text. 

\begin{figure*}[h]
\centering
\includegraphics[width=0.9\textwidth]{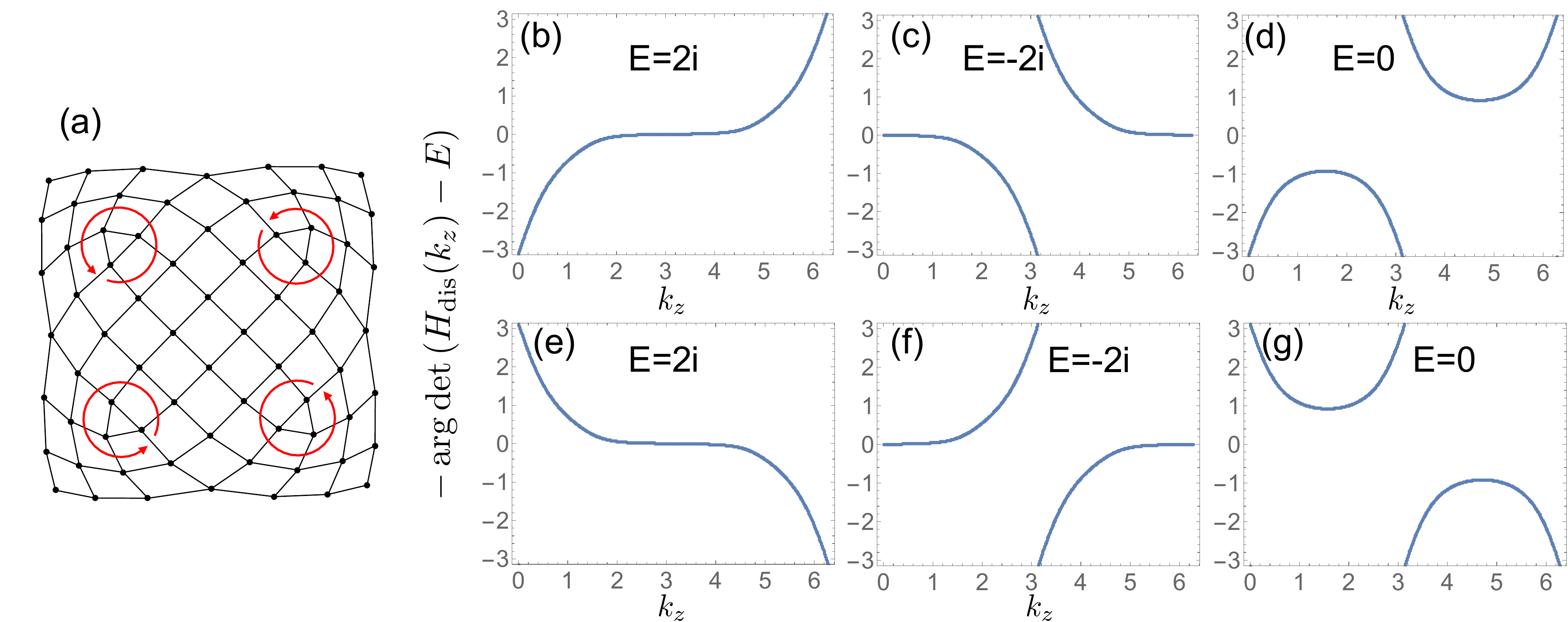}
\caption{Numerical calculations for a non-Hermtian Weyl semimetal described by Eq.~(8) in a lattice where there are four disclinations with Frank angle $\pi/2$ as shown in (a). (b), (c) and (d) are for $\mathcal{L}_{z}=1$. (e), (f) and (g) are for $\mathcal{L}_{z}=-1$.}
\label{fig:minusdis_winding}
\end{figure*}
\begin{figure*}[h]
\centering
\includegraphics[width=1\textwidth]{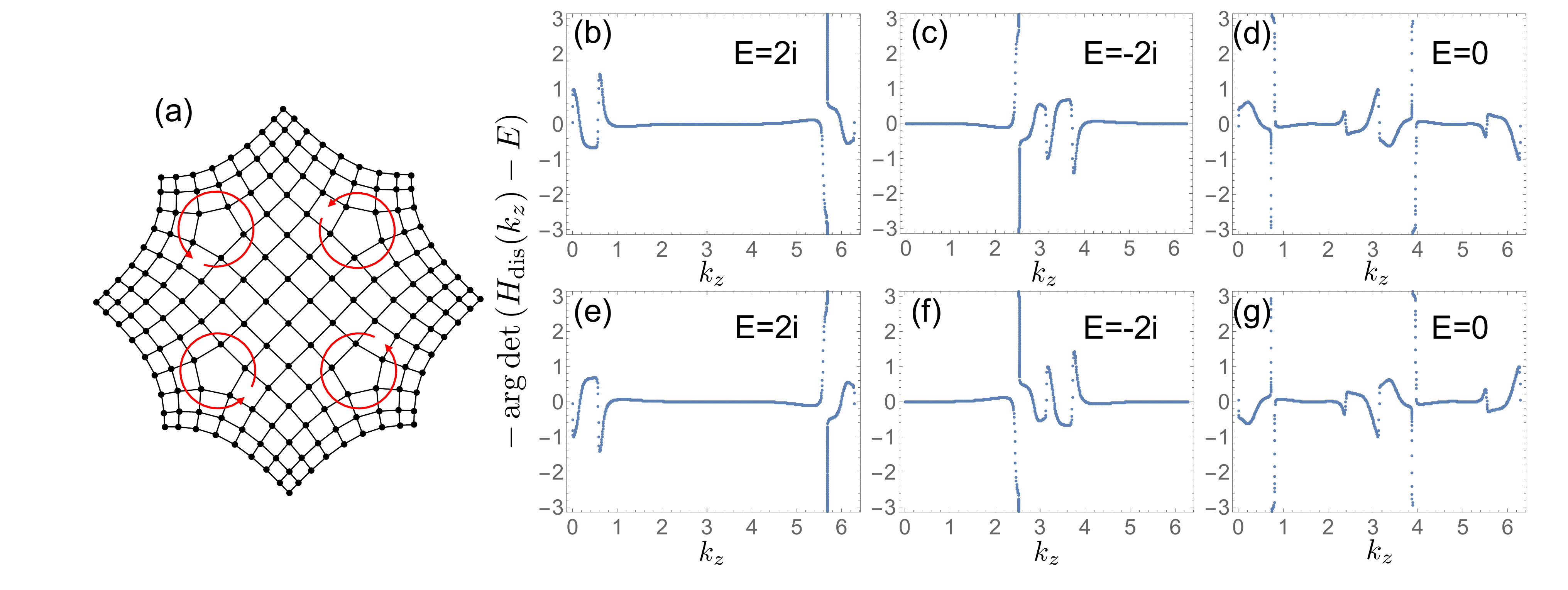}
\caption{Numerical calculations for a non-Hermtian Weyl semimetal described by Eq.~(8) in a lattice where there are four disclinations with Frank angle $-\pi/2$ as shown in (a). (b), (c) and (d) are for $\mathcal{L}_{z}=1$. (e), (f) and (g) are for $\mathcal{L}_{z}=-1$.}
\label{fig:plusdis_winding}
\end{figure*}

\section{Supplemental information for numerical calculations}
Here, we first describe more details about how we perform the numerical calculations. When gluing the open edges at $\phi$ and $\phi+\pi/2$ shown in Fig.~2 (a) in the main text, we add a phase $\exp(-i \mathcal{L}_{z}\pi/2)$. This is because by rotating one edge anti-clockwisely by $\pi/2$ to glue it with another edge, the states on that edge will obtain a phase $\exp(-i \mathcal{L}_{z}\pi/2)$. Meanwhile, we also rotate $\sigma_{x}$ to $\sigma_{y}$, and rotate $\sigma_{y}$ to $-\sigma_{x}$ when gluing. This part is given by the rotation of the internal degrees of freedom (or spin), which is generated by $\sigma_{z}$.

Next, we present numerical results for $\mathcal{L}_{z}=\pm 1$ in the lattice where there are four disclinations, each of them has $\Theta^{s}_{z}=\pm\pi/2$ . In summary, the main conclusions we can draw from these results are: (i) the 1D point gap winding number for the $1<\operatorname{Im}E<3$ case can be correctly predicted by the Wen-Zee term in Eq.~(5). However, (ii) the correct prediction of 1D point gap winding number for the $-3<\operatorname{Im}E<-1$ case and the $-1<\operatorname{Im}E<1$ case requires a combination of rotation and translation lattice information, and can be done through including an extra effective flux as shown in the last section. 

Last, to show that the appearance of disclination-induced skin effect does not depend on the physics of boundaries in $x-y$ directions, we compactify the $x-y$ plane at each $z$ by gluing two lattices with four disclinations, each with Frank angle $\pi/2$, as shown in Fig.~\ref{fig:compact}(a), and then calculate the 1D point gap winding number. Since we have eight ``plaquette-type" disclinations in the lattice, following the index theorem discussed above, we expect to have $W_{1}(E)=2$ when $1<\operatorname{Im}E<3$, $W_{1}(E)=0$ when $-1<\operatorname{Im}E<1$, and $W_{1}(E)=-2$ when $-3<\operatorname{Im}E<-1$, which are confirmed by our numerical results shown in Fig.~\ref{fig:compact}(b-d).

\begin{figure*}[h]
\centering
\includegraphics[width=0.9\textwidth]{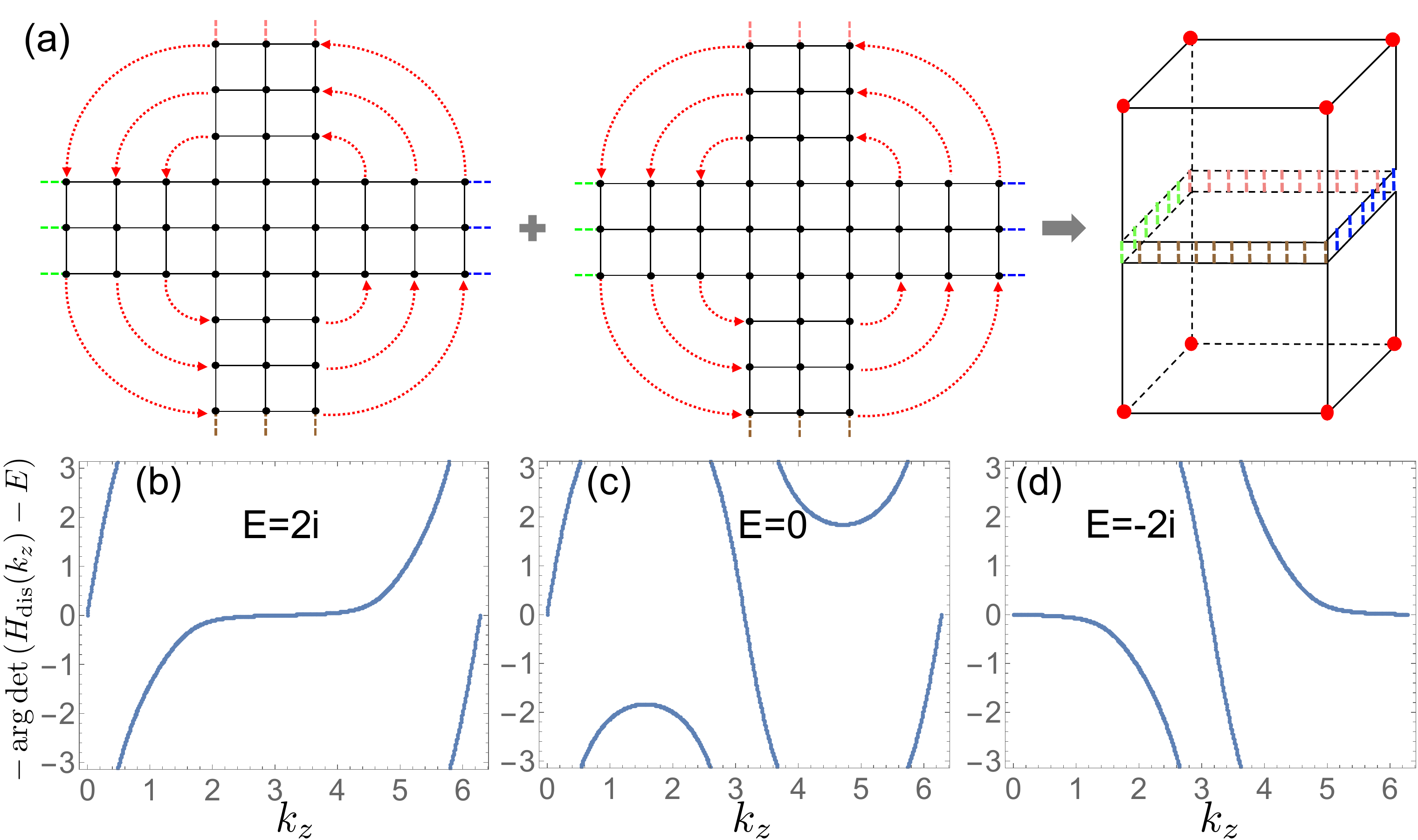}
\caption{(a) illustrates the construction of a lattice with compactified $x-y$ planes and eight disclinations, each with Frank angle $\pi/2$. We take two copies of the disclinated lattice and glue their edges with dotted lines having the same color (pink, green, blue, and brown), then we get the lattice in the rightmost picture, of which the $x-y$ plane at each $z$ is compatified as the surface of a cuboid (that is homeomorphic to a sphere). The eight disclination cores are indicated by the red dots on the right. (b), (c) and (d) shows the 1D point gap winding number for $E=2i$, $E=0$, and $E=-2i$, respectively. }
\label{fig:compact}
\end{figure*}

\end{widetext}

\end{document}